\documentclass[twocolumn,aps,prd,footinbib]{revtex4-1}
\usepackage{comment}
\usepackage{amsmath}
\usepackage{mathtools}
\usepackage{amssymb}
\usepackage{amsfonts}
\usepackage{amsthm}
\usepackage{mathrsfs}
\usepackage[all]{xy}
\usepackage[colorlinks=true, citecolor=blue,linkcolor=blue,urlcolor=blue,linktocpage]{hyperref}
\usepackage{tensor}
\usepackage{footnote}
\usepackage{graphicx} 
\usepackage[justification=raggedright,singlelinecheck=false]{caption}
\usepackage{subcaption}
\usepackage{cleveref}
\usepackage{booktabs}
\usepackage{tabularx}
\newcommand{\ra}[1]{\renewcommand{\arraystretch}{#1}}

\newcommand{\dd}{\mathrm{d}}
\newcommand{\pd}{\partial}

\newcommand{\dx}{\Delta x}
\newcommand{\beq}{\begin{equation}}
\newcommand{\eeq}{\end{equation}}

\def\al{\alpha}
\def\be{\beta}
\def\ga{\gamma}

\def\si{\sigma}

\begin{document}

\title{On the geometry of small causal diamonds}
\author{Jinzhao Wang }
\email{jinzwang@phys.ethz.ch}
\affiliation{Institute for Theoretical Physics, ETH Z\"urich}

\begin{abstract}
The geometry of small causal diamonds is systematically studied, based on three distinct constructions that are common in the literature, namely the geodesic ball, the Alexandrov interval and the lightcone cut. The causal diamond geometry is calculated perturbatively using Riemann normal coordinate expansion up to the leading order in both vacuum and non-vacuum. We provide a collection of results including the area of the codimension-two edge, the maximal hypersurface volume and their isoperimetric ratio for each construction, which will be useful for any applications involving the quantitative properties of causal diamonds. In particular, by solving the dynamical equations of the expansion and the shear on the lightcone, we find that intriguingly only the lightcone cut construction yields an area deficit proportional to the Bel-Robinson superenergy density $W$ in four dimensional spacetime, but such a direct connection fails to hold in any other dimension. We also compute the volume of the Alexandrov interval causal diamond in vacuum, which we believe is important but missing from the literatures. Our work complements and extends the earlier works on the causal diamond geometry by Gibbons and Solodukhin~\cite{gibbons2007geometry},  Jacobson, Senovilla and Speranza~\cite{jacobson2018area} and others~\cite{myrheim1978statistical,roy2013discrete,jubb2017geometry}. Some potential applications of our results in mathematical general relativity and quantum gravity are discussed. 
\end{abstract}

\maketitle

\section{Introduction}

Feynman interpreted the Einstein field equation as directly relating radius excess of some small spatial ball with the matter energy contained within, while holding the area same as in flat Minkowski space~\cite{feynman2018feynman}. It suggests the essence of spacetime dynamics is captured by the geometry of a small causal diamond. Following the same philosophy, there are some proposals demonstrating that the Einstein field equation can be derived from the entanglement equilibrium~\cite{jacobson2016entanglement} or the quantum speed limit~\cite{lloyd2012quantum} using the causal diamond setup. The causal diamond setup also plays an important role in the study of quantum gravity. It has been used in causal set theory~\cite{myrheim1978statistical,buck2015boundary,jubb2017geometry,benincasa2010scalar,roy2013discrete,khetrapal2013boundary}, holography~\cite{de2016entanglement, bousso2002holographic,casini2003geometrical} and cosmology~\cite{bousso2000positive,bousso2007predicting}. In most applications, however, the geometry of the small causal diamond is resolved at the order of Ricci curvature.  It is thus worth studying the geometry at higher order in vacuum to open up more applications of causal diamonds.

Causal diamond is a somewhat broad term and its exact construction differs in different applications. One notable recent work using causal diamond is by Jacobson et al~\cite{jacobson2018area,jacobson2016entanglement}, in which the Einstein field equation is related to the entanglement equilibrium associated with the geodesic ball causal diamond (GCD). The edge area deficit of GCD is governed by the Einstein tensor. By hypothesising that the vacuum entanglement entropy in a small geodesic ball is maximal at fixed volume with respect to variations in both geometry and quantum fields, Jacobson derives the full nonlinear Einstein equation. It is then natural to ask what a higher order perturbation might imply according to the maximal vacuum entanglement hypothesis. One expects at higher order the gravitational superenergy characterised by the Bel-Robinson superenergy $W$ (to be defined later) should be the relevant quantities that controls the GCD geometry.  However, the geometric perturbation does not behave nicely at higher order as shown in~\cite{jacobson2018area}, so ball deformations are considered in~\cite{jacobson2018area} using the Alexandrov interval construction and various other consistent prescriptions. On the other hand, one can take a different perspective by treating the Alexandrov interval as another construction of causal diamonds (ACD) parallel to the geodesic ball construction. In fact, the standard notion of causal diamond usually refers to the Alexandrov interval. ACD already receives a lot of attention in earlier works~\cite{gibbons2007geometry,myrheim1978statistical,roy2013discrete,buck2015boundary}. However, none of the investigations on ACD extends to higher order of perturbations, in particular, the leading order in vacuum. Filling this gap will certainly help extend the various applications of ACD to the vacuum cases, for instance in causal set theory. Furthermore, we also consider a third construction, the lightcone cut, which is used in evaluating the small sphere limit of various quasilocal mass (QLM) proposals~\cite{horowitz1982note,brown1999canonical,chen2018evaluating,kelly1986quasi,bergqvist1994energy,szabados2009quasi}.  One can construct a causal diamond from the lightcone cut (LCD) by taking its domain of dependence. One notable feature of the QLM is that it should give the stress tensor and Bel-Robinson superenergy $W$ in non-vacuum and vacuum respectively at the small sphere limit. It is natural to expect that the geometry of LCD has a nice connection with $W$ in vacuum. We will show, perhaps surprisingly, that among the three constructions, only the lightcone cut construction in four dimensional spacetime yields the result that the edge area deficit being proportional to the superenergy $W$. 

In this work, we unite all three constructions under the same setting of causal diamonds. Since almost all studies of the causal diamond geometry so far restricts to the leading order in the presence of matter, we aim to provide a collection of higher order results that could be of interest for any applications involving the quantitative geometric features of causal diamonds. In order to compute the perturbations of the causal diamond geometry, one could try the `special case' method used in \cite{gibbons2007geometry,jubb2017geometry}, where the causal diamond geometry is evaluated in example spacetimes and universal geometric variations can be drawn out. This method, though convenient, is difficult to generalise to higher order for our purposes. Hence, we adopt the same framework used by \cite{jacobson2018area} in studying GCD to probe the geometry of ACD in Riemann normal coordinates (RNC). Such higher order results are made possible by using Brewin's results of general RNC expansions computed by Cadabra~\cite{brewin2009riemann}. The latter is a powerful tool that allows the small geometries of causal diamonds to be perturbatively probed at arbitrary order of interest. In the case of LCD, as opposed to the method used to investigate the GCD and ACD, we solve the Raychauduri equation and the evolution equation of the shear to evaluate the area of the lightcone cut. We give explicit results for the area of a codimension-2 causal diamond edge $A$, the maximal hypersurface volume bounded by the causal diamond edge $V$ and the respective isoperimetric ratio $I$ between the edge area and the maximal volume. We investigate both non-vacuum and vacuum cases for small ACD and LCD, and relevant GCD geometry calculated in~\cite{jacobson2018area} will also be mentioned for completeness. Furthermore, the d-volume $V^{(d)}$ of the Alexandrov interval causal diamond in vacuum is computed, extending the result in \cite{gibbons2007geometry}. This particular result is missing from the literatures and could have direct applications in the causal set approach to quantum gravity.

In section \ref{sec:pre}, we first introduce some preliminary notions relevant to our discussions, such as the electro-magnetic decomposition of the Weyl tensor, the Bel-Robinson tensor and superenergy density $W$. In section \ref{sec:cds}, we give the three different constructions of causal diamonds. In section \ref{sec:nonvac}, we review and add some results regarding the geometry of small causal diamonds in non-vacuum. Section \ref{sec:vac} contains the new results for the vacuum causal diamonds. We also compute the total volume of ACD in section \ref{sec:acdvol}. Finally in section \ref{sec:discussions}, we briefly discuss the applications of our results.

We use the following index notation: $a,b,c,\dots$ for abstract index notations; $\mu,\nu,\alpha,\dots = 0,1,\dots,d-1$ for RNC expressions concerning the full spacetime; $i,j,k,\dots = 1,\dots,d-1$ for codimension-1 objects and $A,B,C,\dots = 2,\dots,d-1$ for codimension-2 objects. For the sake of brevity, we shall sometimes drop the Big O error term in our results, such as $O(l^{d+2})$, where the expansion order is understood.

\section{Preliminaries}\label{sec:pre}

We are interested in variations of the causal diamond geometry as compared to its flat space counterparts. In non-vacuum, the geometric variations are characterised by Ricci-related quantities, like $R, R_{ab}, G_{ab}$. In vacuum, the Ricci tensor vanishes so the spacetime geometry is characterised by the Weyl tensor $C_{abcd}$. The geometric quantities of interest, such as area and volume, have leading order expansions in terms of the squares of the Weyl tensor, and they can be categorised by the electro-magnetic decomposition of the Weyl tensor. Our discussions on the electro-magnetic decomposition shall only concern relevant notions that we need. One can refer to \cite{senovilla2000super} for more details.

Given some timelike vector $U^a$ at $O$, one can decompose the Weyl tensor at $O$ into spatial tensors on any hypersurface orthogonal to $U^a$ at $O$. These spatial tensors are referred as the electric and magnetic parts.
In adapted coordinates with respect to $U^a$ where the unit normal has coordinates $U^\mu=\delta^\mu_0$, the electric-magnetic decomposition is defined as:
\beq
E_{ij} := C_{0i0j},\;\;\;\;\; H_{ijk} := C_{0ijk} ,\;\;\;\;\; D_{ijkl} := C_{ijkl},
\eeq
where $E_{ij}$ is the electric-electric part, $H_{ijk}$ is the electric-magnetic part  
and $D_{ijkl}$ is the magnetic-magnetic part. 
 The unique tensor with the dominant property\footnote{The dominant property means that the tensor $T_{abcd}$ contracted with any four future directed causal vectors is non-negative.} and quadratic in the Weyl tensor is the Bel-Robinson tensor, defined in arbitrary dimension by~\cite{senovilla2000super}
\begin{align}
T_{abcd} &= C_{aecf} C_b{}^e{}_d{}^f + C_{aedf} C_b{}^e{}_c{}^f 
- \frac{1}{2} g_{ab}  C_{gecf} C^{ge}{}_d{}^f\nonumber\\
&-\frac{1}{2} g_{cd} C_{aegf} C_b{}^{egf}
+\frac{1}{8} g_{ab} g_{cd} C_{efgh}C^{efgh}. \label{eqn:Tarb}
\end{align}

Given a timelike vector $U^a$ at $O$, the associate superenergy density is defined as $W := T_{abcd}U^aU^bU^cU^d$ and it has been shown to characterise various quasilocal masses of the vacuum gravitational field in the small sphere limit~\cite{horowitz1982note,brown1999canonical,chen2018evaluating,kelly1986quasi,bergqvist1994energy,szabados2009quasi}. 
So we can write $W$ as
\beq \label{eqn:W}
W = \frac12\left[E^2+H^2+\frac14 D^2\right],
\eeq
where $E^2=E^{ij}E_{ij}, H^2=H^{ijk}H_{ijk}, D^2=D^{ijkl}D_{ijkl}$ .

In four dimensions, one recovers the original Bel-Robinson (BR) tensor~\cite{szabados2009quasi}
\beq
T_{abcd} = C_{aecf} C_b{}^e{}_d{}^f + *C_{aecf} *C_b{}^e{}_d{}^f,\;\;\;\; (d=4)
\eeq
which is defined in a way similar to how the electromagnetic stress tensor is built from the electromagnetic tensor. The BR tensor in four dimensions enjoys many nice properties, such as being traceless, totally symmetric and divergence-free in vacuum \cite{senovilla2000super}. The superenergy is  
\beq
W=E^2+B^2, \;\;\;\; (d=4)
\eeq
where $B_{ij}:=\frac{1}{2}\epsilon_{jkl}H\indices{_i^k^l}$, and $D^2=4E^2$ when $d=4$.

This form suggests the name `superenergy' density analogous to the field energy in electrodynamics, but with a different dimension. It is in parallel with the fact that in vacuum the BR tensor is divergence-free. In fact, using dimensional analysis, one can argue that in four dimensional vacuum any Lorentz invariant quasilocal mass expression for a small sphere must be proportional to $W$ at leading order \cite{szabados2009quasi}. This justifies the interpretation of $W$ as some gravitational energy. We will investigate three causal diamonds in vacuum and study how the geometric quantities associated with them are related to the electro-magnetic densities $E^2, H^2, D^2$ and particularly the superenergy $W$. 

We close this section by introducing some important expressions and identities that will be useful later in the calculations. The metric expansion in RNC up to the curvature squared order~\cite{brewin2009riemann} is given in terms of the Riemann curvature tensor at the RNC origin $O$:

\begin{align}
g_{\al\be}(x) &= \eta_{\al\be}-\frac13 x^\mu x^\nu R_{\al\mu\be\nu} 
-\frac16 x^\mu x^\nu x^\rho\nabla_\mu R_{\al\nu\be\rho}\nonumber\\
&+ x^\mu x^\nu x^\rho x^\si\left(\frac2{45} R\indices{^\ga_\mu_\al_\nu} R_{\ga\rho\be\si}
-\frac1{20} \nabla_\mu \nabla_\nu
R_{\al\rho\be\si}\right). \label{eqn:metric}
\end{align}

The volume and area integral involves integration over solid angles and the following result~\cite{othmani2011polynomial} will be handy to use in our calculations later: 
\begin{align}
\int_{S^d} \dd \Omega_{d} n^{i_1}\dots n^{i_k}&= \frac{\Omega_d(d-1)!!}{(d+k-1)!!}\delta^{(k)}_{i_1i_2...i_k},  \mbox{($k$ even)}\\
\int_{S^d} \dd \Omega_{d} n^{i_1}\dots n^{i_k}&= 0. \hspace{3cm} \mbox{($k$ odd)}\label{eqn:oddn}
\end{align} \label{eqn:integralsolidangle}
 where $n^i(\theta)$ is the unit spatial vector with $n^in_i=1$, $\{\theta_A\}$ is the angular coordinates on the $d$-sphere $S^d$, $\Omega_d=\frac{2\pi^{(d+1)/2}}{\Gamma(\frac{n+1}{2})}$ is the volume of unit d-sphere $S^d$ and $\delta^{(k)}_{i_1i_2...i_k}$ is defined recursively for even $k$: 
\begin{align}
 \delta^{(k+2)}_{i_1i_2...i_{k+2}}&=(k+1)!\delta_{i(j}^{(2)}\delta^{(k)}_{i_1i_2...i_k)},\nonumber\\
 &=\delta_{ij}\delta^{(k)}_{i_1i_2...i_k}+\delta_{ii_1}\delta^{(k)}_{ji_2...i_k}+\dots+\delta_{ii_k}\delta^{(k)}_{i_1i_2...j}.
\end{align}

 $\delta^{(2)}_{ij}$ is the usual kronecker delta $\delta_{ij}$ and the second equality above is due to the fact that $\delta^{(k)}_{i_1i_2...i_k}$ so defined is totally symmetric. We will frequently use the following in later calculations.
\beq
\delta^{(4)}_{klmn}=\delta_{kl}\delta_{mn}+\delta_{km}\delta_{ln}+\delta_{kn}\delta_{ml}.
\eeq

In vacuum, we also have the following properties concerning the electro-magnetic decompositions of the Weyl tensor. 
\begin{subequations}\label{eqn:identities}
\begin{align}
&E\indices{_i^i}=H\indices{^i_j_i}=0,\\
&D\indices{_{ik}_j^k}=-C\indices{_{i0}_j^0}=C\indices{_{i0j0}}=E_{ij},\\
&H\indices{^k^l_i}H_{lkj}\delta^{ij}=\frac{H^2}{2},\\
&H\indices{^k_i^l}H^{min}\delta^{(4)}_{klmn}=\frac{3}{2}H^2,\\
&E^{kl}E^{mn}\delta^{(4)}_{klmn}=2E^2,\\
&D\indices{_p^k_i^l}D^{pmin}\delta^{(4)}_{klmn}=\frac{3}{2}D^2+E^2,\\
&\delta^{ij}\delta^{(4)klmn}\nabla_k \nabla_l C_{imjn}=0.
\end{align}
\end{subequations}

\section{Three constructions of causal diamonds} \label{sec:cds}

The standard causal diamond, the Alexandrov interval causal diamond, is the intersection region of two lightcones, with one oriented upwards (future-directed) and the other downwards (past-directed). The joint of intersection has the topology of $S^{d-2}$, and we shall refer them as the diamond edge. One can generalise the standard definition by starting from a given edge $S$ and define the domain of dependence as the causal diamond. This is motivated by the fact that it is the edge that captures the subtle geometric features of the causal diamond. In this work, three distinct constructions of the diamond edge found in the literatures are considered. Starting from a closed, spacelike, codimension-2 surface $S$ that is homeomorphic to $S^{d-2}$, and an arbitrary spacelike hypersurface $\Sigma$ with $\partial\Sigma=S$, we define the causal diamond $D_S$ as the domain of dependence $D_S:=D(\Sigma)$. Note that $D(\Sigma)$ is independent of what $\Sigma$ we choose as long as it is spacelike and has the edge $S$ as boundary. The causal diamond defined in this way coincides with the notion of causally closed/complete set in algebraic quantum field theory \cite{casini2003geometrical,haag1996local}, and also resembles the entanglement wedge in AdS/CFT~\cite{rangamani2017holographic}. This is also how a geodesic ball causal diamond is defined in \cite{jacobson2016entanglement}. Therefore, we believe our definition in terms of the domain of depedence associated with the edge is more general than the usual notion of causal diamond as the Alexandrov interval, thus facilitating more potential applications. Our definition also resembles the light-sheet construction by Raphael Bousso's covariant entropy bound~\cite{bousso2002holographic}. His construction is more sophisticated as $S$ can be some arbitrary codimension-2 surface whereas we only consider the closed surface that has a topology of sphere.  

We shall investigate three constructions of the diamond edge, and they give the geodesic ball causal diamond (GCD), the Alexandrov interval causal diamond (ACD), and the lightcone cut causal diamond (LCD). Each causal diamond has an associated orientation and we shall use a normalised timelike vector $U^a\in T_O M$ to characterise it, where $O$ is the reference point associated with the diamond. We will later set up RNC at $O$ and thus call it the diamond origin. It is located in the centre of GCD and ACD, whereas it sits at the lower tip in LCD.

\begin{figure}[t]
  \centering
  \begin{subfigure}[b]{0.3\linewidth}
    \includegraphics[width=\linewidth]{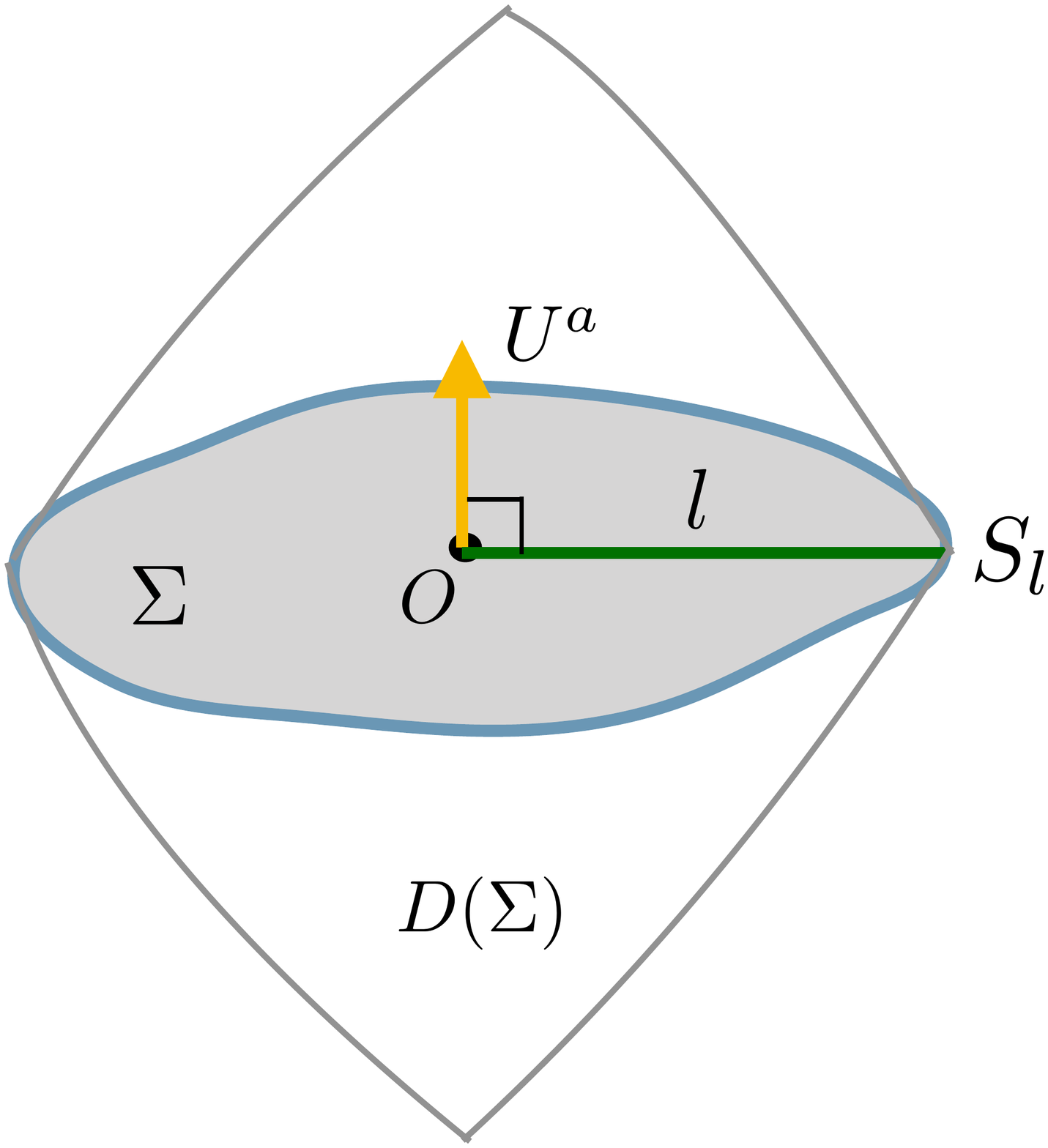}
    \caption{GCD}
    \label{fig:gcd}
  \end{subfigure}
  \begin{subfigure}[b]{0.3\linewidth}
    \includegraphics[width=\linewidth]{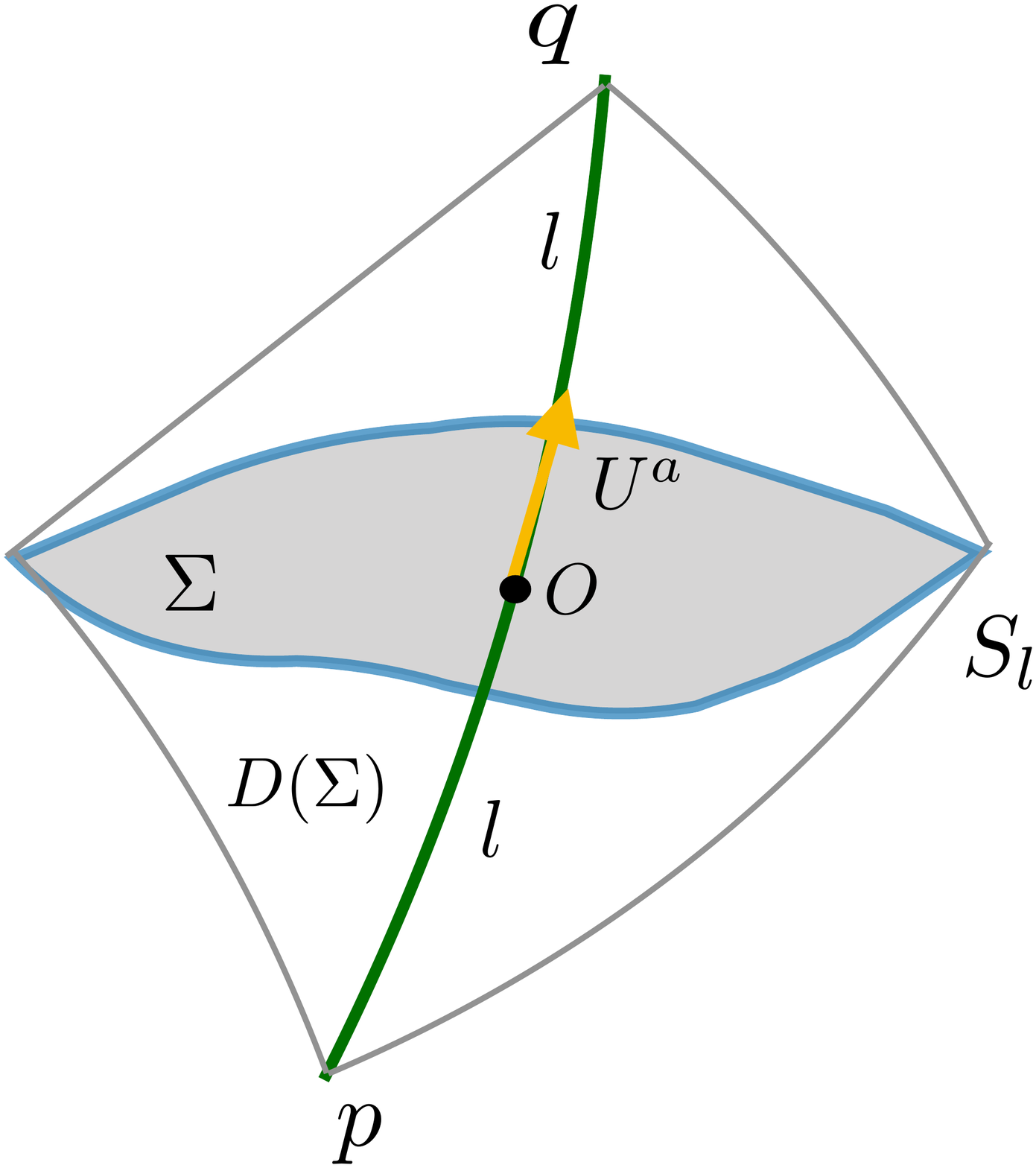}
    \caption{ACD}
    \label{fig:acd}
  \end{subfigure}
   \begin{subfigure}[b]{0.3\linewidth}
    \includegraphics[width=\linewidth]{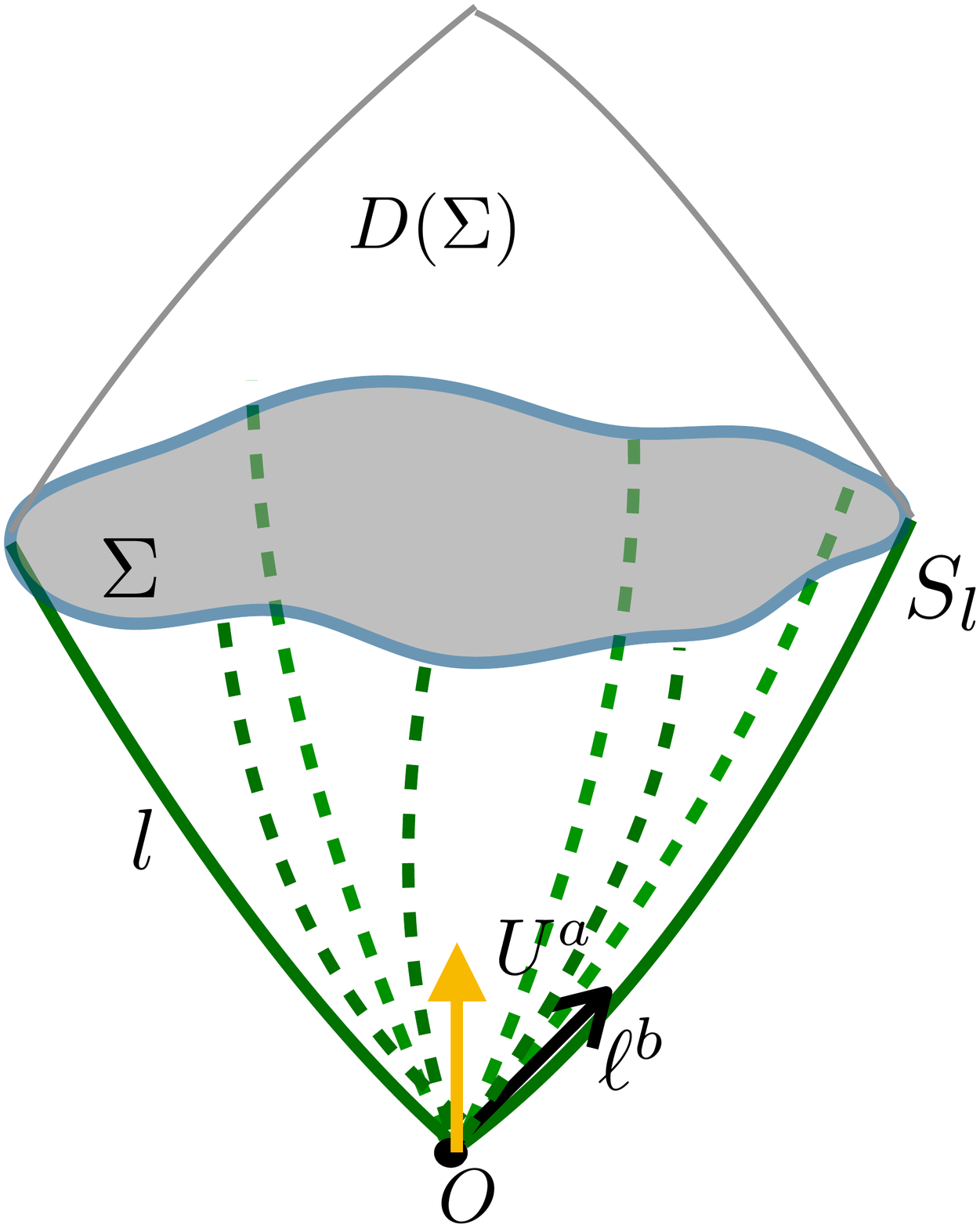}
    \caption{LCD}
    \label{fig:lcd}
  \end{subfigure}
  \caption{Three causal diamond constructions are illustrated here. The size parameter $l$ (green), orientation $U^a$ (yellow), diamond edge $S_l$ (blue) and diamond origin $O$ are indicated for each causal diamond. $\Sigma$ specifies the spacelike hypersurface with maximal volume and its domain of dependence $D(\Sigma)$ defines the causal diamond $D_{S_l}$.} 
\end{figure}

The causal diamonds $D_{S}$ are defined from following edge constructions (See Figure 1 for illustrations):

\textbf{GCD}: The edge in GCD is defined as the $(d-1)$-ball with geodesic radius $l$, with the geodesics emanating from $O$ and orthogonal to $U^a$. Its boundary is the edge $S_l$, where the subscript denotes the size parameter $l$. 

\textbf{ACD}: The edge in ACD is defined with a timelike geodesic interval $\gamma(p,q)$ from $p$ to $q$ with the affine parameter running from $-l$ to $l$. $O$ is the midpoint of $\gamma(p,q)$ and $U^a$ is the unit tangent to the geodesic at $O$. The edge $S_l$ is the intersection of the two lightcones $S_l=\dot{I}^+(p)\cap\dot{I}^-(q)$ and the resulting causal diamond is thus the Alexandrov interval $I^+(p)\cap I^-(q)$, where $I^\pm(\cdot)$ denotes the chronological future/past of a point and $\dot{I}^\pm(\cdot)$ denotes its boundary. 

\textbf{LCD}: Lastly, the edge in LCD is defined as the lightcone cut. More precisely, given $O$ and $U^a$, the future-directed null vector $\ell^a$ is normalised as $U_a\ell^a=-1$ and the level set of parameter distance $l$ along the null rays defines a lightcone cut $S_l$ and it serves as the edge of LCD. 

Note that apart from the origin $O$ and the orientation $U^a$, each diamond has a size parameter $l$, which refers to distinct quantities in the three constructions. Nevertheless, we denote the edge as $S_l$ in all three constructions to keep the notation consistent. Also note that all three constructions are identical in Minkowski spacetime, and we obtain the same causal diamond if the size parameters $l$'s are set equal \footnote{This is why we choose the ACD diamond to have the geodesic interval of length $2l$ as opposed to $l$ found in most literatures.}. Therefore, the way we defined the size parameter for each construction is indeed consistent. One can think of $l$ being small and the order counting in perturbative expressions throughout is based on the power of $l$. Also, we will only study small causal diamonds, where the size $l$ is much smaller than the curvature scale, such that we can do calculations using the RNC expansions and we do not need to worry about caustics or conjugate points.

The above causal diamonds are used by people in different contexts. The GCD is the one used by Feynman to interpret Einstein equation and recently used by Jacobson to derive Einstein equation from entanglement equilibrium~\cite{jacobson2018area,jacobson2016entanglement}. The ACD is perhaps the most standard causal diamond and it is a natural object to consider in causal set theory~\cite{myrheim1978statistical,buck2015boundary,jubb2017geometry,benincasa2010scalar,roy2013discrete,khetrapal2013boundary}. Its geometric properties are investigated in different contexts by Gibbons and Solodukhin~\cite{gibbons2007geometry,gibbons2007geometry2,berthiere2015comparison}. LCD or more precisely the lightcone cut itself \footnote{The lightcone cut in the literatures is usually used without constructing a causal diamond on top of it. Here, we think of it as the edge of LCD in order to keep in accordance with the other two constructions.} is the standard construction of the small sphere in various quasilocal mass proposals~\cite{horowitz1982note,brown1999canonical,chen2018evaluating,kelly1986quasi,bergqvist1994energy,szabados2009quasi}, and its qualitative features are studied in terms of the lightcone comparison theorem by Choquet-Bruhat et al~\cite{choquet2009light}.

We'll be mainly interested in the geometry related to the diamond edge, namely, the $(d-2)$-volume of the edge that we henceforth refer as area $A$, the maximal hypersurface volume $V$ bounded by the edge and the isoperimetric ratio $I$ between them. We will calculate their variations with respect to the causal diamond of the same size in Minkowski spacetime. We use the notation 
\beq\label{eqn:flatparts}
V^\flat=\frac{\Omega_{d-2}l^{d-1}}{d-1} ,\;\; A^\flat=\Omega_{d-2}l^{d-2}
\eeq
to denote the maximal hypersurface volume and edge area in Minkowski spacetime. One probably wonder whether the different recipes for causal diamonds actually differ in the perturbative order that we are interested in. The answer is definitive and the distinction is more manifested at leading order in vacuum. Nevertheless, one can already partially distinguish them in presence of matter as we shall now discuss.

\section{The non-vacuum case} \label{sec:nonvac}
In presence of matter, the earlier studies investigated area/volume deficits of causal diamonds, in particular GCD and ACD. The relevant quantities that govern the geometry of small diamonds are Ricci related quantities defined at the diamond origin $O$, where the $0$-component refers to the basis aligning with the orientation $U^a$ of the diamond. Here we review some known results and add some new results concerning LCD. We shall consider causal diamonds of some varied size $l+X_m$ where $X_m$ is the size ambiguity with respect to the size $l$ in Minkowski spacetime. More explanation and elaboration will be provided later in context. We assume the Einstein field equation $G_{ab}=8\pi GT_{ab}$ throughout this section.

\subsection{Geodesic ball Causal Diamond}
Jacobson showed \cite{jacobson2016entanglement} that for a small GCD with radius $l$ and orientation $U^a$, the volume and area deficits are
\begin{align}
\delta V &=-\frac{\Omega_{d-2}l^{d+1}G_{00}}{3(d-1)(d+1)}=-\frac{8\pi G\Omega_{d-2}l^{d+1}}{3(d-1)(d+1)}T_{00},\label{eqn:gcdnonvacuum}\\
\delta A &=-\frac{\Omega_{d-2}l^d G_{00}}{3(d-1)}=-\frac{8\pi G\Omega_{d-2}l^d}{3(d-1)}T_{00},
\end{align}
where $G_{00}$ is the component of the stress energy tensor along the orientation of the diamond $G_{ab}U^aU^b$, and similarly for other such quantities throughout the paper.
Therefore, the Einstein equation can be viewed as an equation relating the area/volume deficit of GCD edge with the stress energy density. This is how Feynman interpreted the Einstein equation in section 11.2 of \cite{feynman2018feynman} \footnote{To be more precise, Feynman actually fixes the area and considers the radius surplus.}, and is also part of Jacobson's argument in deriving Einstein equation from the condition of maximal entanglement entropy associated with GCD.

We can compute the isoperimetric ratio between the area and volume given by the following formula
\beq\label{eqn:isoGCD1}
I:=\frac{V/V^\flat}{(A/A^\flat)^{\frac{d-1}{d-2}}}=1+\frac{G_{00}l^2}{(d-2)(d+1)}+ O(l^3).
\eeq
The isoperimetric ratio is defined such that $I=1$ in flat spacetime, and we see that its variation is also proportional to $G_{00}$.

\subsection{Alexandrov Interval Causal Diamond}
ACD is the most commonly used causal diamond. It serves as a very useful and natural setup in causal set theory. The geometry of small ACD has been studied by Gibbons and Solodukhin~\cite{gibbons2007geometry}. For an Alexandrov interval of length $2l$, we have the variations of maximal hypersurface volume and edge area at leading order:
\begin{align}
\delta V &=-\frac{\Omega_{d-2}l^{d+1}(R-(d-1)R_{00})}{6(d^2-1)} + O(l^{d+2}), \label{eqn:ACDvol1}\\
\delta A &=-\frac{\Omega_{d-2}l^d(R-(d-4)R_{00})}{6(d-1)}+ O(l^{d+1}). \label{eqn:ACDarea1}
\end{align}

How is ACD different with GCD at this order? In fact, one can relax the GCD radius from being fixed and consider deformations of the geodesic ball in GCD and make it ACD. Take the radius in GCD to be $r=l+\frac{R_{00}}{6(d-1)}l^3$, and one will find that they match exactly with $\delta V, \delta A$ above (\ref{eqn:ACDvol1},\ref{eqn:ACDarea1}). Therefore, one can think the ACD as the scaled GCD. This match is only possible at this order, i.e. the leading order in non-vacuum, and we shall see later that at higher order, the edge in ACD also deviates along the $U^a$ direction away from the geodesic ball orthogonal to $U^a$, making the two causal diamonds incomparable. 

Motivated by the above considerations, one can in general relax fixing the size parameter $l$ with respect to the Minkowski diamond, but rather introduce a size ambiguity as $l+X_m$, where the subscript $m$ stands for matter. This is also done in \cite{jacobson2018area} when considering more general variations of GCD. We can set $X_m\sim O(l^3)$ such that the dimensionless variations due to the size ambiguity $X_m/l \sim O(l^2)$ is of the same order as the variations due to curvature $R_{00}l^2, Rl^2, G_{00}l^2$. Here, we take the simplification that $X_m$ does not have angular dependence. Since $X_m/l$ is chosen at the same order as the leading order of our interest, there is actually no loss of generality by assuming $X_m$ to be spherically symmetric, as the angular dependence will average out after integration eventually. Accommodating $X_m$ gives us the flexibility to fix any other geometric quantities, like area, volume or more sophisticated ones. We can thus extend all the results by appending an ambiguity variation term $X_m$ to the curvature variations, and we will see the usefulness of explicitly stating the ambiguity in some examples later. Nevertheless, if one is only interested in variations when the size parameter is fixed, the size ambiguity $X_m$ can be set as zero. Later in discussions of vacuum case, a similar ambiguity at higher order $X_v$ is introduced as well.

Let us take the example of ACD just mentioned, $X_m$ is an ambiguity on the proper length between two vertices of the diamond. The variations $\delta V', \delta A'$ due to $X_m$ can be readily computed from the flat space values
\begin{align}
\delta V' &=V^\flat(l+X_m)-V^\flat(l)=\Omega_{d-2}l^{d-2}X_m ,\\
\delta A' &=A^\flat(l+X_m)-A^\flat(l)=\Omega_{d-2}l^{d-3}(d-2)X_m.
\end{align}

Appending these to (\ref{eqn:ACDvol1},\ref{eqn:ACDarea1}): 
\begin{align}
\delta V &=-\frac{\Omega_{d-2}l^{d+1}(R-(d-1)R_{00})}{6(d^2-1)}+\Omega_{d-2}l^{d-2}X_m,\label{eqn:ACDvol1.1}\\ 
\delta A &=-\frac{\Omega_{d-2}l^d(R-(d-4)R_{00})}{6(d-1)}+\Omega_{d-2}l^{d-3}(d-2)X_m. \label{eqn:ACDarea1.1}
\end{align}

Taking $X_m=-\frac{R_{00}}{6(d-1)}l^3$ gives the GCD results. Also notice the minus sign here. This is the reverse of turning GCD to ACD as pointed out above. 

We see that the area/volume deficits cannot both be proportional to $T_{00}$ in ACD with any choice of $X_m$ , but it turns out that the surplus of isoperimetric ratio $I$ between them does~\cite{gibbons2007geometry}.
\beq
I:=\frac{V/V^\flat}{(A/A^\flat)^{\frac{d-1}{d-2}}}=1+\frac{G_{00}l^2}{(d-2)(d+1)} =1+\frac{8\pi GT_{00}l^2}{(d-2)(d+1)}.
\eeq
The isoperimetric ratio is independent of an overall change in the size of the causal diamond, so $X_m$ is not manifested in the variation above. This is therefore not surprising that this isoperimetric ratio is the same as the one of GCD in (\ref{eqn:isoGCD1}). 

\subsection{Lightcone cut Causal Diamond}
The geometry of small LCD in arbitrary dimension has not been systematically studied even in non-vacuum, except for that, qualitatively, comparison theorems regarding the area of the cut has also been given in~\cite{choquet2009light}. Here we intend to compute the edge area and maximal hypersurface volume of the LCD. In addition, we also give a recipe for associate a non-vanishing volume form to the lightcone itself and compute its volume under this prescription.

Recall that the construction of the lightcone cut. One starts with $\ell^a, U^a\in T_O(M)$, where $U^a$ is the timelike vector which defines the orientation of the causal diamond, $\ell^a$ is the null generator of the lightcone. The affine parameter $l$ of the null congruence is normalised by imposing $U_a \ell^a=-1$. We choose the basis of $T_O(M)$ such that $U^\mu=(1,0,0,0), \ell^\mu=(1,n^i), n^in_i=1$. The RNC is set up around the tip of lightcone $O$ and the coordinates of a generic point $p$ on the lightcone are $x^\mu(p)=(l,ln^i)$, where $l$ is the parameter distance between $p$ and $O$. The lightcone cut which defines the edge $S_l$ of LCD is the locus of points with the affine parameter distance $l$ on the lightcone. Hence, $l$ controls the size of the causal diamond and we shall calculate the relevant quantities up to order $O(l^4)$. 

\subsubsection{Edge area}
By solving the Raychaudhuri equation and the evolution equation of the shear, the edge area deficit in small LCD is shown to be
\beq \label{eqn:LCDareanonvac}
\delta A=-\frac{\Omega_{d-2}l^d(dR_{00}+R)}{6(d-1)}+\Omega_{d-2}l^{d-3}(d-2)X_m+ O(l^{d+1}).
\eeq
We defer the calculation details to section \ref{sec:lcdvacarea} where we solve the equations to higher orders of perturbations that cover both vacuum and non-vacuum cases.

\subsubsection{Maximal hypersurface volume}

We can also make an attempt to evaluate the maximal hypersurface volume enclosed by the lightcone cut $S_l$. In Minkowski spacetime, the maximal slice corresponds to the geodesic ball with radius $l$. We expect that the perturbed maximal hypersurface deviates from the geodesic ball. One can treat the lightcone cut as the corrugated boundary of some deformed geodesic ball. In order to have a good control of the deviation, we shall assume that the causal diamond with orientation $U^a$ has spherical symmetry. With this assumption, the maximal slice is either a geodesic ball orthogonal to the geodesic generated by $U^a$ or a cone-shaped spacelike hypersurface with a conical singularity. The latter cannot have maximal volume, so we end up with a ball that deviates from the flat ball only radially.

To evaluate the deviated geodesic ball volume, we first need to locate the deformed geodesic ball with the lightcone cut as its boundary. We do so by changing from the RNC centred at the lightcone vertex $O$ to another RNC centred at $O'$, which is $l'$ parameter distance away following the geodesic generated by $U^a$. We use the following transformation formula between RNCs to the leading order \cite{brewin2009riemann}.
\beq
x'^\mu(p) = \Delta x^\mu + \frac13 R\indices{^\mu_{\alpha\nu\beta}}\;x^\nu_{O'}\Delta x^\alpha \Delta x^\beta.
\eeq
where $x'^\mu(p), x^\mu(p)$ denote the coordinates of a generic event $p$ in $O'$-RNC and $O$-RNC respectively; $x^\mu_{O'}$ is the coordinates of event $O'$ in $O$-RNC; and $\Delta x^\mu=x^\mu(p)-x^\mu_{O'}$. We've left out a possible local Lorentz transform in the tangent space, which is not relevant for our coordinate transform.

Take $p$ to be some point in the lightcone cut, it has $x^\mu(p)=(l,l n^i)$. We also have $x_{O'}^\mu=(l',\mathbf{0})$. Since we should choose $l'$ in such a way that the ball deviates from the flat ball only radially, and the radial geodesics are orthogonal to $U^a$, so in RNC we should set $x'^0(p)=0.$ By substitution to the above equation, we have
\beq \label{eqn:LCDmaxslice}
x'^k(p) = l n^k+ \frac13 R\indices{_0_i^k_j}n^in^j l'l^2-\frac13 R\indices{_0^k_0_i}n^i(l-l')l'l
\eeq
with the constraint
\beq
x'^0(p) = l-l' - \frac13 R_{0i0j}n^in^jl'l^2 \equiv 0.
\eeq
The above constraint gives $l'=l- \frac13 R_{0i0j}n^in^jl^3+O(l^4)$. Plugging it into (\ref{eqn:LCDmaxslice}) gives
\begin{align}
x'^k(p) =& l n^k+\frac{1}{3}  R\indices{_0_i^k_j}n^in^j l^3-\frac19 R\indices{_0^k_0_i}R\indices{_0_j_0_l}n^in^jn^l l^5\nonumber\\
&-\frac19 R\indices{_0_i^k_j}R_{0m0n}n^in^jn^mn^n l^5.
\end{align}

This above expression does not capture a radial deviation yet. Because of spherical symmetry, we can set $x'^k(p)=rn'^k$ with some radius $r$. We can fix $r$ by taking the inner product on both sides of the above expression:
\beq
r^2=l^2+O(l^6),
\eeq
where we've taken the average over $\Omega_{d-2}$, so we have
\beq
x'^k(p)=ln'^k.
\eeq
Hence, the non-trivial term is beyond the order of interest, so the conclusion is that the geodesic ball is effectively not deviated at leading order in vacuum and we have the ball boundary sitting at $x'(p)=(0,l n'^k)$. Its volume deficit is purely due to the metric change and is given by (\ref{eqn:gcdnonvacuum}). However, later we see that these perturbation terms do contribute in vacuum. Hence, we have the maximal hypersurface volume deficit of a spherical symmetric LCD
\beq\label{eqn:LCDmaxvolnonva}
\delta V=-\frac{\Omega_{d-2}l^{d+1}G_{00}}{3(d-1)(d+1)}+\Omega_{d-2}l^{d-2}X_m+ O(l^{d+2}).
\eeq
It is less meaningful to compute the isoperimetric ratio for LCD here as the above volume is only valid assuming spherical symmetry. Hence, we leave this out for future works.

Motivated by~\cite{jacobson2016entanglement}, there is a good reason to fix the volume rather than the size parameter. We can check if we can connect the volume deficit to $T_{00}$ by holding volume constant rather than the size parameter $l$. This is the case for GCD, and thus ACD as argued above. 
Holding volume fixed corresponds to a variation in the size parameter, by solving $\delta V(X_m)=0$, we have
\beq\label{eqn:fixingvolLCD}
X_m = - \frac{\delta V|_l}{\Omega_{d-2}l^{d-2}},
\eeq
where $V|_l$ means holding size $l$ fixed, i.e. $V|_l=-\frac{8\pi G\Omega_{d-2}l^{d+1}}{3(d-1)(d+1)}T_{00}.$
Now using (\ref{eqn:LCDareanonvac}), we can readily compute the area holding the volume fixed:
\begin{align}
\delta A|_V&=\delta A|_l +\Omega_{d-2}l^{d-3}(d-2)X_m=\delta A|_l - \frac{d-2}{l}\delta V|_l, \nonumber\\
&=-\frac{\Omega_{d-2}l^d(d^2-d+4)R_{00}+3R}{6(d^2-1)},
\end{align}
which unfortunately fails to connect directly with the stress energy tensor.

We hereby summarise and collect all the results mentioned and computed in this section into Table \ref{tab:nonvac} below. The first row is the Minkowski reference of the geometric quantities, which is the same for all three diamond constructions. The other entries are the corresponding variations to the flat space values. We've set $X_m=0$ to keep the table clean.

\begin{table*}[ht]
    \centering
    \ra{1.9}
    \begin{ruledtabular}\begin{tabularx}{\textwidth}{@{}cccc@{}}
        \midrule
       Background &Edge area $A$ &Maximal hypersurface volume $V$ &Isoperimetric ratio $I$ \\ \toprule \hline
        $\mathbb{M}^d$  &$\Omega_{d-2}l^{d-2}$ &$\frac{\Omega_{d-2}l^{d-1}}{d-1}$ & $1$
        \\
        GCD &$-\frac{\Omega_{d-2}l^dG_{00}}{3(d-1)}$ & $-\frac{\Omega_{d-2}l^{d+1}G_{00}}{3(d-1)(d+1)}$ & $\frac{G_{00}l^2}{(d-2)(d+1)}$
        \\
        ACD &$-\frac{\Omega_{d-2}l^d(R+(d-4)R_{00})}{6(d-1)}$ &$-\frac{\Omega_{d-2}l^{d+1}(R+(d-1)R_{00})}{6(d^2-1)}$ &$\frac{G_{00}l^2}{(d-2)(d+1)}$
        \\
        LCD &$-\frac{\Omega_{d-2}l^d(dR_{00}+R)}{6(d-1)}$  &$-\frac{\Omega_{d-2}l^{d+1}G_{00}}{3(d-1)(d+1)}$ & /%$\frac{(d^2-d+4)R_{00}+3R}{6(d-2)(d+1)}l^2$
        \\
         \bottomrule
    \end{tabularx}\end{ruledtabular}
    \caption{\textbf{The leading order geometry of small causal diamonds in non-vacuum.} The first row shows the geometry in flat Minkowski background and the rows below give the leading order deviations of the edge area, the maximal hypersurface volume and the isoperimetric ratio for each diamond in non-vacuum.}\label{tab:nonvac}
\end{table*}

%%%%%%%%%%%%%%%%%%%%%%%%%%%%%%

\section{The vacuum case}\label{sec:vac}

We shall now work with Ricci-flat spacetime and we are interested in the order of fourth derivative of the metric. In vacuum, the Riemann tensor equals to the Weyl tensor and we will use $R_{abcd}$ and $C_{abcd}$ interchangeably. As in the non-vacuum case, we introduce a size ambiguity $l+X_v$ to the size parameter in all three constructions, where the subscript $v$ stands for vacuum. We set $X_v \sim O(l^5)$, such that the dimensionless perturbation $X_v/l\sim O(l^4)$ is at the same order as other perturbative quantities characterised by $E^2l^4, H^2l^4, D^2l^4$ due to curvature. $X_v$ introduces the flexibility to fix any other geometric quantity of interest besides the size parameter at the same order. In the following we will evaluate areas and volumes with the size fixed, and then append the variation due to $X_v$ in the end. Nevertheless, if one is only interested in variations when the size is fixed, the size ambiguity $X_v$ can be set as zero.\\

\subsection{Geodesic ball Causal Diamond}
GCD has been studied by Jacobson et al in~\cite{jacobson2018area}, so we simply summarise their results here. The hypersurface volume variation is
\beq \label{eqn:gcdvoldeficit}
\delta V=\frac{\Omega_{d-2} l^{d+3}}{15(d^2-1)(d+3)}\left[ -\frac{D^2}{8} - \frac{H^2}{2} +\frac{E^2}{3}\right]+\Omega_{d-2}l^{d-2}X_v,
\eeq
and the edge area deficit is 
\beq\label{eqn:gcdareadeficit}
\delta A = \frac{\Omega_{d-2} l^{d+2}}{15(d^2-1)}\left[-\frac{D^2}{8} -\frac{H^2}{2} +\frac{E^2}{3}\right]+\Omega_{d-2}l^{d-3}(d-2)X_v.
\eeq

As for the non-vacuum results, we've appended the size ambiguity terms accordingly. We see that as opposed to the GCD in non-vacuum, here the variations are not always negative definite and thus not proportional to $W$ as given in (\ref{eqn:W}). Jacobson et al also considered various plausible deformations of the geodesic ball motivated from different perspectives. The deformation considered coincides with the prescription of ACD and the leading order deformation is specified as $\delta r=\frac{1}{6}l^3n^in^jE_{ij}$. The second order variation is unspecified and denoted as $X$ in \cite{jacobson2018area} and our investigation of ACD in the next subsection shall fill in this gap. Lastly, the isoperimetric ratio surplus can be readily computed
\beq
I=1+\frac{\left(3D^2 +12H^2 -8E^2\right)l^4}{72(d+3)(d+1)(d-2)}+O(l^5).
\eeq

\subsection{Alexandrov Interval Causal Diamond}
In ACD, our task is to first fix the coordinates of the edge. Since the two vertices $p,q$ of ACD are given, we treat it as a geodesic boundary value problem. By putting the constraints of the vanishing arc-length of the null geodesics sitting at the lightcone, we can solve for the coordinates of the edge in RNC base at the centre $O$ of the diamond. The arc-length of a generic geodesic interval starting at $x^\mu$ and ending at $x^\mu+\dx^\mu$ in RNC~\cite{brewin2009riemann} has the following expression up to the order of interest:
\begin{widetext}
\begin{align}\label{eqn:arc}
L^2=& \eta_{\mu\nu}\dx^\mu\dx^\nu-\frac{1}{3}R_{\mu\alpha\nu\beta}x^\alpha x^\beta \dx^\mu\dx^\nu-\frac{1}{12}(\nabla_\mu R_{\nu\beta\alpha\rho}+2\nabla_\alpha R_{\mu\beta\nu\rho})x^\alpha x^\beta x^\rho \dx^\mu\dx^\nu\nonumber \\
-&\Big(\frac{1}{45}R_{\mu\rho\nu\sigma}R_{\alpha\lambda\beta}^{\;\;\;\;\;\;\;\sigma}+\frac{1}{60}\nabla_\mu\nabla_\nu R_{\alpha\rho\beta\lambda}\Big) x^\rho x^\lambda \dx^\mu\dx^\nu\dx^\alpha\dx^\beta+\Big(\frac{2}{45}R_{\mu\alpha\beta\sigma}R_{\nu\rho\lambda}^{\;\;\;\;\;\;\;\sigma}-\frac{1}{20}\nabla_\alpha\nabla_\beta R_{\mu\rho\nu\lambda}\Big)x^\alpha x^\beta x^\rho x^\lambda \dx^\mu\dx^\nu\nonumber \\
+&\Big(\frac{2}{45}R_{\mu\alpha\beta\sigma}R_{\nu\rho\lambda}^{\;\;\;\;\;\;\;\sigma}-\frac{1}{20}\nabla_{(\mu}\nabla_{\beta)} R_{\nu\rho\alpha\lambda}\Big)x^\beta x^\rho x^\lambda \dx^\mu\dx^\nu\dx^\alpha .
\end{align}
\end{widetext}

RNC is set up around the midpoint $O$ of the geodesic interval $\gamma(p,q)$. We choose basis of $T_O(M)$ such that time direction is set to be the orientation of the diamond. Hence $q, p$ have coordinates $(\pm l,0,0,0)$ respectively, and the edge $S_l$ is located at $x^\mu_S=(t, r(\theta)n^i)$ where $n^i$ is normalised $n^in_i=1$. $t(\theta), r(\theta)$ are unknown coordinate components describing the time and radial directions respectively, which depends on the angular coordinates $\{\theta_A\}$. We have 
\beq
x_{q,p}^\mu=(\pm l,0,0,0), \;\;\dx^\nu = x^\mu_S-x^\mu_{q,p}=(t(\theta)\mp l,r(\theta)n^i),
\eeq

and we can now plug these into (\ref{eqn:arc}) and set $L^2=0$. This gives us two equations and each corresponds to null generators emanating from the top $q$, and bottom $p$ of the causal diamond. Solving them simultaneously for $t(\theta), r(\theta)$ yields the equations that describe the edge.

The solutions are:
\begin{widetext}
\begin{align}
r(\theta)&=l+\frac{1}{6}E_{ij}n^in^jl^3+\frac{1}{24}n\cdot\nabla E_{ij}n^in^jl^4\nonumber\\
&+\frac1{120}l^5\Big[\frac{1}{3}\Big((11E_{ij}E_{lk}+4H\indices{_i^m_j}
H_{lmk})n^in^jn^ln^k-8E_{il}H\indices{_j^l_k}n^in^jn^k+4E\indices{_i^l}E_{lj}n^in^j\Big)\nonumber \\ 
&+2n^k\nabla_k\nabla_0E_{ij}n^in^j +\nabla_0\nabla_0E_{ij}n^in^j+n^ln^k\nabla_l\nabla_k E_{ij}n^in^j\Big] + O(l^6) ,\label{eqn:radius}\\
t(\theta)&=-\frac{1}{24}\nabla_0 E_{ij}n^in^j l^4+\Big(\frac{1}{45}E_{il}H\indices{_j^l_k}+\frac{1}{40}\nabla_k\nabla_0 E_{ij}\Big)n^in^jn^kl^5 + O(l^6) .\label{eqn:time}
\end{align}
\end{widetext}

We see that at the leading order in non-vacuum $O(l^3)$, there are radial deviations from flat geometry but no temporal ones, which only kicks in when considering leading order in vacuum $O(l^5)$. This timelike deviation (\ref{eqn:time}) complicates our evaluation of the edge area and the volume of the maximal hypersurface. Nevertheless, we can circumvent the complications at this perturbative order. Since the geodesic ball orthogonal to $U^a$ maximises the spatial volume in flat space, we can assume that the maximal hypersurface in ACD is of the form that perturbs the geodesic ball. The variations of the geometric quantities of interest associated with the perturbed ball will be characterised by the $l^3, l^4, l^5$ terms in (\ref{eqn:radius},\ref{eqn:time}). In particular, the contributions due to the $l^4, l^5$ can be calculated already by averaging over the solid angles, because any combinations of these two terms with other perturbative terms will have order higher than $O(l^5)$. Note that in (\ref{eqn:time}) the $l^4$ term average to zero in vacuum and the $l^5$ term will vanish according to (\ref{eqn:oddn}).

Therefore, we can safely ignore these perturbations in the time direction, and the ACD at this order is thus effectively equivalent to GCD with a radius variation. Note that at leading order in non-vacuum, the ACD can be converted to deformed GCD exactly and one can think of them being equivalent up to scaling, but here they are only effectively equivalent in terms of those integral quantities of interest, namely the volume and the area. After averaging, (\ref{eqn:radius}) simplifies to

\beq\label{eqn:radiusdeformation}
r(\theta)=l+\frac16 E_{ij}n^in^jl^3+\frac{(2d+13)E^2+3H^2}{180(d^2-1)}l^5 + O(l^6).
\eeq

We've argued that the ACD edge can be effectively treated as a deformed geodesic ball. The induced metric $h_{ij}$ on the ball is 
\begin{align}
h_{ij} (x) =&\; \delta_{ij} -\frac13 x^k x^l R_{ikjl} - \frac16 x^k x^l x^m \nabla_k R_{iljm} \nonumber \\
&\;+ x^k x^l x^m x^n\Big(-\frac2{45} R_{0kil}R_{0mjn}  \nonumber \\
&+ \frac2{45} R\indices{^p_k_i_l}R_{pmjn} 
-\frac1{20} \nabla_k\nabla_l R_{imjn} \Big )+O(x^5) .\label{eqn:hij}
\end{align} 

To compute the edge area, we need the pullback metric on the edge.
\begin{align}
q_{AB} = \frac{\pd x^\mu}{\pd\theta_A}\Big|_{S_l}\frac{\pd x^\nu}{\pd\theta_B}\Big|_{S_l}g_{\mu\nu}\Big|_{S_l}=\frac{\pd r(\theta)n^i}{\pd\theta_A}\frac{\pd r(\theta)n^j}{\pd\theta_B}h_{ij}.\label{eqn:qab}
\end{align}
where $r(\theta)$ is given by (\ref{eqn:radiusdeformation})

The maximal hypersurface volume and the edge area integrals in spherical coordinates are given by
\begin{align}
V &= \int\sqrt{h} d^{d-1} x = \int d\Omega_{d-2} \int_0^{r(\theta)} dr\, r^{d-2} \sqrt{h}, \label{eqn:vol}\\
A &=\int \sqrt{q} \dd^{d-2}\theta = l^{d-2} \int_{S_l} \dd\Omega_{d-2} \sqrt{q}, \label{eqn:area}
\end{align}

\begin{widetext}
The integrals are calculated in~\cite{jacobson2018area}, and we simply quote their results. 
\begin{align}
V &=V^\flat (l) +\Delta V+ \Omega_{d-2} l^{d-3}  \left[l X +\frac{(d-2)Y^{ij}Y_{ij}}{d^2-1} -\frac{l^3}{3(d^2-1)} Y^{ij}E_{ij}  \right] +O(l^{d+4}),\label{eqn:totalV}\\
A &=A^{\flat}(l) +\Delta A + \Omega_{d-2} l^{d-4}\left[(d-2)l X +\frac{d^2-3d+4}{d^2-1}Y^{ij}Y_{ij}-\frac{l^3 d}{3(d^2-1)} Y^{ij} E_{ij} \right]+O(l^{d+3}), \label{eqn:totalA}
\end{align}
with $\delta V, \Delta A$ given by (\ref{eqn:gcdvoldeficit}, \ref{eqn:gcdareadeficit}). $Y_{ij},X$ are radius deformation $r=l+Y_{ij}n^in^j+X$. 

Compared with (\ref{eqn:radiusdeformation}), we can substitute $$Y_{ij}=\frac{l^3}6E_{ij},\; X=\frac{(2d+13)E^2+3H^2}{180(d^2-1)}l^5,$$
into (\ref{eqn:totalV},\ref{eqn:totalA}) to obtain the variations 
\begin{align}
\delta V &=\Omega_{d-2}l^{d+3}\frac{(14d^2+28d-34)E^2+6(d+1)H^2-3D^2}{360(d-1)(d+1)(d+3)}+\Omega_{d-2}l^{d-3}(d-2)X_v+O(l^{d+4}), \\
\delta A &=\Omega_{d-2}l^{d+2}\frac{(14d^2-32d-4)E^2+6(d-4)H^2-3D^2}{360(d-1)(d+1)}+\Omega_{d-2}l^{d-2}X_v+O(l^{d+3}).
\end{align}
\end{widetext}
Note that here the size ambiguity $X_v$ here has nothing to do with $X$. The latter was treated as a size ambiguity of GCD in \cite{jacobson2018area} but fixed by ACD geometry here. We see that including the higher order variation $X$, as suggested in \cite{jacobson2018area}, does not help make the area deficit proportional to $W$. Another interesting quantity to consider is the isoperimetric ratio $I$. As discussed for the non-vacuum ACD, $\delta I$ is proportional to the stress-energy tensor at leading order. It is therefore plausible that $\delta I$ in vacuum is proportional to the superenergy at leading order.
\begin{align}
I := &\frac{V/V^\flat}{(A/A^\flat)^{\frac{d-1}{d-2}}},\nonumber\\
=& 1+\frac{\left((-2d^2+2d+16)E^2+12H^2+3D^2\right)l^4}{72(d+3)(d+1)(d-2)},\nonumber\\
= &1+\frac{\left(12W-(d+1)(d-2)E^2\right)l^4}{36(d+3)(d+1)(d-2)},
\end{align}
where in the last line we have substituted in $W$ according to (\ref{eqn:W}). Unfortunately, this ratio variation fails to directly connect with the $W$ unlike in the non-vacuum case. 
%\newpage
\subsection{Lightcone cut Causal Diamond}

We first compute the area of the cut in vacuum for arbitrary spacetime dimension $d$, which has not been investigated before to our knowledge. 

\subsubsection{Area of the lightcone cut}\label{sec:lcdvacarea}
There are two ways of computing the area. The first entails solving evolution equations of the optical quantities on the lightcone. It uses the dynamical behaviors of the lightcone and is less demanding in terms of calculations, so we are going to follow this method. The second method is conceptually simpler, only demanding the pullback metric on $S_l$ in RNC and then the area integral, but the calculation is much lengthier \footnote{The direct calculation of the lightcone cut area is demonstrated in section A of Supplemental Material at [URL will be inserted by publisher].}. Since we shall also demonstrate the area in non-vacuum case (\ref{eqn:LCDareanonvac}) with the same method, we will not assume Ricci tensor vanishes a priori in this subsection. 

Denote the covariant derivative of the null generators as $B_{ab}:=\nabla_a\ell_b$. One can decompose $B_{ab}$ into the twist, expansion and shear. The lightcone has vanishing twist. The expansion and shear are defined by the following:
\begin{align}
\hat{\theta} &:= \hat{B}^a_a= \nabla_a \ell^a,\\
\hat{\sigma}_{ab} &:=\hat{B}_{(ab)} - \frac{1}{d-2}\hat{\theta} h_{ab},
\end{align}
where $\hat{\cdot}$ denotes the transverse projection given by 
\beq
h^a_b = \delta^a_b + N^a\ell_b + \ell^aN_b,
\eeq
where $N^a$ is a null vector field on the lightcone obeying:
\beq
N^a\ell_a = -1,\;\;\nabla_\ell N^a = 0.
\eeq
In RNC, following the LCD setup outlined in earlier sections ($\ell^\mu=(1,n^i), U^\mu=(1,0,0,0)$), we have
\beq\label{eqn:projectionRNC}
N^\mu=(1/2,-n^i/2), h^\mu_\nu=\delta_j^\mu\delta^i_\nu(\delta^i_j-n^in_j),
\eeq
where the repeated $i,j$ here are not summed. Hence in RNC $h^\mu_\nu$ is block-diagonalised with $\mathbf{0}\oplus (\delta^i_j-n^in_j)$, so it only projects to the spatial part as expected \footnote{Note that we've only used the leading order expression for $h^\mu_\nu$ and $N^\mu$ as the higher order contributions will be irrelevant for our calculations later. }.

The expansion of the null geodesic congruence governs the rate of change of the lightcone cut area. Denote the pullback metric on the cut $S_l$ to be $q_{AB}$, and its volume form satisfies:
\beq\label{eqn:areachange}
\dot{\sqrt{q}} = \theta\sqrt{q},
\eeq
where the dot represents the derivative with respect to the affine parameter $l$ of the null generators.\\
With vanishing twist on the lightcone, the Raychaudhuri equation and the evolution equation for shear in arbitrary dimension $d$ read \cite{burger2018towards}:
\begin{align}
\dot{\hat{\theta}} &= -\frac{1}{d-2}\hat{\theta}^2 - \hat{\sigma}^{ab} \hat{\sigma}_{ab} - R_{ab}\ell^a\ell^b,\label{eqn:raychaudhuri1}\\
\dot{\hat{\sigma}}_{ab} &= -\frac{2}{d-2}\hat{\theta}\hat{\sigma}_{ab}-C_{cedf} h^c_a \ell^e h^d_b \ell^f ,\label{eqn:raychaudhuri2}
\end{align}
where $C_{cedf}$ is the Weyl tensor and we shall use a short hand for the partly projected Weyl term 
\beq
C_{ab}:=C_{cedf} h^c_a \ell^e h^d_b \ell^f.
\eeq
Note that the Weyl tensor appeared in the shear evolution equation should be defined at $x^\mu=l\ell^\mu$. Nevertheless, we still use the tensor $C_{cedf}$ evaluated at the origin, because the difference between the Weyl tensor at $l$ and origin can be safely ignored for our leading order calcualtions later. In RNC, the non-zero components of $C_{ab}$ can be computed using (\ref{eqn:projectionRNC}). Because of the projection, only the spatial parts $C_{ij}$ are non-zero.
\begin{align}
C_{ij} =& E_{ij} - 2n^kE_{k(i}n_{j)}+n_in_jE_{lk}n^ln^k-2H_{(ij)k}n^k\nonumber\\
&+2n_{(i}H\indices{^l_j_)^k}n_ln_k+D_{ikjl}n^kn^l.
\end{align}

Since we are only interested in perturbative solutions to (\ref{eqn:raychaudhuri1},\ref{eqn:raychaudhuri2}), the ODEs can be solved by a power series ansatz. It is known that the lightcone expansion is given by $\hat{\theta}=(d-2)/l$ in Minkowski spacetime. Substituting it into (\ref{eqn:raychaudhuri2}) yields
$\hat{\sigma}_{ab}=-C_{ab}l/3$ at the leading order. Therefore, we propose the following ansatz:
\begin{align}
\hat{\theta}(l)& = \frac{d-2}{l} + c_0 + c_1 l + c_2 l^2 + c_3 l^3 +O(l^4),\\
\hat{\sigma}_{ab}(l)& = -\frac{C_{ab}}{3} l + k_2 l^2 + k_3 l^3 +O(l^4).
\end{align}
Plugging into the (\ref{eqn:raychaudhuri1},\ref{eqn:raychaudhuri2}) and solving them simultaneously up to $O(l^3)$ gives:
\begin{align}\label{eqn:expansion}
\hat{\theta}(l)& = \frac{d-2}{l} -\frac{R_{ab}\ell^a\ell^b}{3} l - \frac{(d-2)C_{ab}C^{ab}+(R_{ab}\ell^a\ell^b)^2}{45(d-2)} l^3,\\
\hat{\sigma}_{ab}(l)& = -\frac{C_{ab}}{3} l - \frac{2 C_{ab}R_{cd}\ell^c\ell^d}{45(d-2)} l^3.
\end{align}

We are only interested in the expansion to compute the volume form $\sqrt{q}$. We shall once again use an ansatz:
\beq
\sqrt{q} = \Omega_{d-2}l^{d-2}\left(1+q_1 l + q_2 l^2 + q_3 l^3 + q_4 l^4 \right) + O(l^{d+3}).
\eeq
Plugging in the ansatz and (\ref{eqn:expansion}) into (\ref{eqn:areachange}) yields:
\begin{align}
\sqrt{q} &= \Omega_{d-2}l^{d-2}\Big(1-\frac{R_{ab}\ell^a\ell^b}{6} l^2 \nonumber\\
&- \frac{(2d-4)C_{ab}C^{ab}+(12-5d)(R_{ab}\ell^a\ell^b)^2}{360(d-2)} l^4. \label{eqn:areaelement}
\end{align}

Now we are ready to compute the edge area. We first show (\ref{eqn:LCDareanonvac}), and we only need the first two terms in (\ref{eqn:areaelement}). 
\begin{align}
\int_{S_l} \sqrt{q}\dd x^{d-2}&=l^{d-2}\int_{S_l} \dd \Omega_{d-2} \left( 1-\frac{R_{ab}\ell^a\ell^b}{6}l^2\right),\nonumber\\
&=l^{d-2}\Omega_{d-2}\left(1-\frac{(d-1)R_{00}+R\indices{^i_i}}{6(d-1)}\right),\nonumber\\
&=A^\flat\left( 1-\frac{dR_{00}+R}{6(d-1)}l^2\right) + O(l^{d+1}).
\end{align}
which gives the result (\ref{eqn:LCDareanonvac}). 

In vacuum, $R_{ab}=0$ and (\ref{eqn:areaelement}) reduces to 
\beq
\sqrt{q} = \Omega_{d-2}l^{d-2}\left(1 - \frac{C_{ab}C^{ab}}{180} l^4 \right) + O(l^{d+3}).
\eeq
Computing $C_{ab}C^{ab}$ is lengthy and state the result here:
\begin{align}
C_{ab}C^{ab}&=E^2-2E\indices{_i^k}E\indices{^k_j}n^in^j+2E_{ij}D\indices{^i_l^j_k}n^ln^k\nonumber\\
&+E_{ij}E_{lk}n^in^jn^ln^k+4H\indices{^k^l_i}H_{(kl)j}n^in^j\nonumber\\
&-2H\indices{_i^m_j}H_{lmk}n^in^jn^ln^k\nonumber\\
&+D\indices{^m_i^n_j}D_{mlnk}n^in^jn^ln^k.
\end{align}
Finally, the integration yields:
\begin{align}
A&=l^{d-2}\int_{S_l} \dd \Omega_{d-2} \left( 1- \frac{C_{ab}C^{ab}}{180} l^4\right),\nonumber\\
&=A^\flat\left(1-\frac{\left(2(d^2+2)E^2+3D^2+6dH^2\right)l^4}{360(d^2-1)}\right),
\end{align}

which in particular when $d=4$, with $D^2=4E^2, H^2=2B^2$, it gives
\beq
\delta A=-\frac{2 l^6\Omega_{2}}{225}(E^2+B^2)=-\frac{2l^6\Omega_2}{225}W.
\eeq
whereas it is not proportional to $W$ in any other spacetime dimension.

The negative definite area variation can be understood using the lightcone cut comparison theorem in the case of the energy condition being satisfied trivially~\cite{choquet2009light}. This area deficit in four dimensions was partially implied by the volume form mentioned in studies of quasilocal mass~\cite{horowitz1982note,brown1999canonical,kelly1986quasi,bergqvist1994energy}. Their calculations are carried out in the Newman-Penrose formalism and therefore only restricted to $d=4$. We here show that this nice connection between area deficit and LCD surprisingly holds only in dimension four. \\

If we relax the identification of the size parameter $l$ with reference to the Minkowski spacetime, introducing a size ambiguity $X_v$ gives the the general result of the lightcone cut area deficit
\begin{align}
\delta A=-\Omega_{d-2}l^{d+2}\frac{2(d^2+2)E^2+3D^2+6dH^2}{360(d^2-1)}\nonumber\\
+\Omega_{d-2}l^{d-3}(d-2)X_v+O(l^{d+3}).
\end{align}

\subsubsection{Maximal hypersurface volume}

The maximal hypersurface volume is straightforward to evaluate following exactly the same procedure given in the non-vacuum case. We state the results here and the details can be found in \footnote{The maximal hypersurface volume of LCD in vacuum is calculated in section B of Supplemental Material at [URL will be inserted by publisher].}. The volume variation assuming spherical symmetry is
\begin{align}
\delta V =& \frac{\Omega_{d-2}l^{d+3}\left(-(40d+112)E^2+(d+2)12H^2-3D^2\right)}{360(d^2-1)(d+3)}\nonumber\\
&+\Omega_{d-2}l^{d-2}X_v+O(l^{d+4}).
\end{align}

Again, we will not compute isoperimetric ratio of LCD here as the above hypersurface volume is only valid under spherical symmetry. We close this section by summarising all the results in the table below. We see that none of the quantities have a direct connection with $W$ (\ref{eqn:W}) in all dimensions. To keep the expressions as simple as possible, we set the size ambiguity $X_v$ to be zero.
\begin{table*}[ht]
    \centering
    \ra{1.9}
   \begin{ruledtabular} \begin{tabularx}{1.1\textwidth}{@{}cccc@{}}
        \toprule
       Background &Edge area $A$ &Maximal hypersurface volume $V$ &Isoperimetric ratio $I$ \\ 
       \hline
        $\mathbb{M}^d$  &$\Omega_{d-2}l^{d-2}$ &$\frac{\Omega_{d-2}l^{d-1}}{d-1}$ & $1$
        \\
        GCD &$\Omega_{d-2} l^{d+2}\frac{\left( -\frac{D^2}{8} - \frac{H^2}{2} +\frac{E^2}{3}\right)}{15(d^2-1)}$ & $\Omega_{d-2} l^{d+3}\frac{\left( -\frac{D^2}{8} - \frac{H^2}{2} +\frac{E^2}{3}\right)}{15(d^2-1)(d+3)}$ & $\frac{\left(\frac{D^2}{8} +\frac{H^2}{2} -\frac{E^2}{3}\right)l^4}{3(d+3)(d+1)(d-2)}$
        \\
        ACD &$\Omega_{d-2}l^{d+2}\frac{(14d^2-32d-4)E^2+6(d-4)H^2-3D^2}{360(d-1)(d+1)}$ &$\Omega_{d-2}l^{d+3}\frac{(14d^2+28d-34)E^2+6(d+1)H^2-3D^2}{360(d-1)(d+1)(d+3)}$ &$\frac{\left((-2d^2+2d+16)E^2+12H^2+3D^2\right)l^4}{72(d+3)(d+1)(d-2)}$
        \\
        LCD &$-\Omega_{d-2}l^{d+2}\frac{\left(2(d^2+2)E^2+3D^2+6dH^2\right)}{360(d^2-1)}$ &$-\Omega_{d-2}l^{d+3}\frac{\left((40d+112)E^2-12(d+2)H^2+3D^2\right)}{360(d^2-1)(d+3)}$  & /
        \\
         \bottomrule
  \end{tabularx}  \end{ruledtabular}
    \caption{{\bf The leading order geometry of small causal diamonds in vacuum.} The first row shows the geometry in flat Minkowski background and the rows below give the leading order deviations of the edge area, the maximal hypersurface volume and the isoperimetric ratio for each diamond in vacuum.}\label{tab:vac}
\end{table*}

\section{Volume of ACD in vacuum}\label{sec:acdvol}

The total volume of ACD has been computed in \cite{gibbons2007geometry}  up to the leading order in non-vacuum. Since the volume variation of ACD provides a link between the continum geometric quantities like Ricci scalar curvature and the discrete counting of k-chains in causal set theory~\cite{roy2013discrete}, it is worth working out the volume expansion to leading order in vacuum. The result could also be used to test the discrete causal set action for an ACD region~\cite{buck2015boundary}.

The $d$-volume integral can be expressed as the sum of the upper cone volume and lower cone volume:
\begin{align}
V^{(d)}&= \int \dd \Omega_{d-2}\int^l_{t(\theta)}\dd t\int^{r^+(t,\theta)}_0\sqrt{g}\; r^{d-2}\dd r \nonumber\\
&+ \int \dd \Omega_{d-2}\int_{-l}^{t(\theta)}\dd t\int^{r^-(t,\theta)}_0\sqrt{g}\; r^{d-2}\dd r.
\end{align}
where $t(\theta)$ locates the diamond edge as given by (\ref{eqn:time}). Via imposing vanishing (\ref{eqn:arc}) on the lightcone, similar to how (\ref{eqn:radius},\ref{eqn:time}) are obtained, the lightcone boundary $r^\pm(t,\theta)$ of ACD is given by the following equation:
\begin{widetext}
\begin{align}
r^\pm(t,\theta)=(l\mp t)\left\lbrace1+\frac{E_{ij}n^in^j}{6}l^2+\frac{E_{ij}E_{lk}n^in^jn^ln^k}{24}l^4\mp\frac{E\indices{_i^k}E_{kj}n^in^j}{45}l^3t + \frac{l^2(l\mp t)^2n^in^j}{90}\left(E\indices{_i^l}E_{lj}+\left[H\indices{_i^m_j}H_{lmk}-E_{ij}E_{lk}\right]n^ln^k\right)\right\rbrace,
\end{align}
\end{widetext}
where $+$ indicates the upper cone and $-$ indicates the lower cone and they simply differ by the sign of $t$. Note that the above expression is the abridged version, where we omit terms that will be irrelevant in the integration later according to (\ref{eqn:identities}). One can find the full expression in \footnote{The calculation details of ACD volume is in section C of Supplemental Material at [URL will be inserted by publisher].}. As a sanity check, one sees that at the edge $r^\pm(0,n)=t(\theta)$ agrees with (\ref{eqn:radiusdeformation}) as expected.

Following similar argument in calculating the maximal hypersurface volume of ACD, only the averaged $t(\theta)$, which is zero, will be relevant. Therefore, we can simplify the integrals, and the upper and lower cones contribute the same 
\beq
V^{(d)}= 2\int \dd \Omega_{d-2}\int^l_0\dd t\int^{r^+(t,\theta)}_0\dd r\; r^{d-2}\sqrt{g},
\eeq
with 
\begin{widetext}
\begin{align}
\sqrt{g} =& 1-\frac{1}{180}C\indices{^\ga_\mu^\al_\nu} C_{\ga\rho\al\si}x^\mu x^\nu x^\rho x^\si, \\
=&1-\frac{1}{180}\Big[E^2t^4 +(2E^{lk}D_{likj}+4H\indices{^l_i^k}H\indices{_{(l|j|k)}}+2E_{ik}E\indices{^k_j})n^in^jr^2t^2+r^4(E_{ij}E_{lk}-2H\indices{_i^m_j}H_{lmk}+D\indices{^m_i^n_j}D\indices{_{mlnk}})n^in^jn^ln^k \Big]. \label{eqn:determinantvariation}
\end{align}
Again, we've keep only the relevant terms. Note that the leading order of above expansion is already of order $R^2$, so we can divide the integral into two parts.

\begin{align}
V^{(d)} = & -\frac{1}{90}\int \dd \Omega_{d-2}\int^l_0\dd t\int^{l-t}_0\dd r\; r^{d-2} \delta \sqrt{g}+2\int \dd \Omega_{d-2}\int^l_0\dd t\int^{r^+(t,\theta)}_0\dd r\; r^{d-2},
\end{align}
where we've kept the first term of $r^+=l-t$ in the first integral and $\delta\sqrt{g}$ is given by (\ref{eqn:determinantvariation}) above.\\

The final result is
\beq
V^{(d)}=\frac{2\Omega_{d-2}l^d}{d(d-1)}+\Omega_{d-2}l^{d+4}\frac{(7d^3+58d^2+146d+108)E^2+6(d+2)(d+6)H^2-3(d+2)D^2/2}{90 (d-1) (d+1) (d+2) (d+3) (d+4)}+O(l^{d+5})
\eeq

where the first term is the flat space d-volume and the second term is the variation due to curvature.\\
\end{widetext}

\section{Discussion} \label{sec:discussions}

The leading order causal diamond geometry is investigated perturbatively in our work, both in vacuum and non-vacuum. It systematically complements and extends the earlier investigations in \cite{gibbons2007geometry,jacobson2018area,myrheim1978statistical,roy2013discrete,jubb2017geometry}. We've summarised our results in Table \ref{tab:nonvac} \& \ref{tab:vac} above, and hopefully this glossary will be useful to those who work with causal diamonds. There are still a few missing pieces from our results. We're not aware of a general technique to determine the maximal surface given some arbitrary closed boundary. It would be interesting to find a way of doing this and then lift the spherical symmetry assumption in our analysis of the maximal hypersurface volume in LCD. One may also be interested in the non-vacuum expansions up to the same order as we explored in the vacuum case. We've left them out, but they can be done simply by keeping all those Ricci terms that are set to zero in vacuum. In principle, following the same method outlined in our work, one can compute the geometry of small causal diamond up to arbitrary order of interest. For that, one needs higher order RNC expansions. Doing this by hand is a daunting task, but fortunately they can be computed by a powerful tool, Cadabra, following the guidelines provided by Leo Brewin~\cite{brewin2009riemann}. Nevertheless, we can hardly think of any cases where a higher order result will be useful. Furthermore, one can apply the same methods to probe the geometry of small causal cones, which are constructed by intersecting a light cone with a spacelike hypersurface. This is partly done in \cite{jubb2017geometry} and one can work out higher order geometries following the same strategy as we tackle causal diamonds. 

We do not attempt to interpret our results here, rather we would like to discuss some potential applications of small causal diamonds. Since the three causal diamond constructions have a wide range of applications in studying gravitational theory, our discussion here doesn't mean to be comprehensive. One can also refer to the discussions in \cite{gibbons2007geometry,jacobson2018area}.

Following the theme of \cite{jacobson2018area}, we want to look for a causal diamond construction that yields the area deficit proportional the Bel-Robinson Superenergy density $W$ in vacuum. LCD turns out to be the only construction that gives such nice relation exclusively in dimension four. However, LCD is not so nice that the same relation with the stress tensor fails in non-vacuum. It is still possible that one can directly connect area deficit with $T_{00}$ or $W$, with the size ambiguity $X_{m}$ determined by holding some quantities fixed in any of the three causal diamonds, or even with another different recipe for causal diamond. 

Since LCD is commonly used in the small sphere limit of quasilocal mass, it would be interesting to see whether the connection between various QLM proposals still obey the small sphere limit in arbitrary dimensions.  Provided the QLM proposals could admit higher dimensional generalisations, such as the generalised Hawking mass and the Brown-York mass as proposed in \cite{miao2017quasi}, one can apply the same techniques used in our calculations to such proposals to verify the small sphere limits. Since the area deficit hints at four dimensions being somewhat unique, it could be that the small sphere limit of QLM's also fails to be proportional to $W$ in vacuum in dimensions other than four.

We fixed the ACD geometry by solving the geodesic boundary value problem for the lightcones, and then computed the integrals. Our methods of probing ACD can be applied to causal set theory. For example, ACD is used to test the boundary term contribution in the causal set action~\cite{buck2015boundary}. It turns out that in Minkowski spacetime, the BDG causal set action~\cite{benincasa2010scalar} evaluated in ACD contributes an amount proportional to the edge area. It would be interesting to check, using the same machinery as in section \ref{sec:acdvol}, if the same holds true in general vacuum spacetime. Another case where ACD is used is in calculating the discrete Ricci curvature and Ricci scalar in terms of counting of k-chains in a causal set sprinkled from a small ACD~\cite{roy2013discrete}. One can generalise their results to higher order,  to obtain the casual set counterparts of other geometric quantities like the electro-magnetic decompositions of the Weyl tensor $E^2, H^2, D^2$ and the Bel-Robinson superenergy density $W$.

There are other plausible applications of our result. Causal diamonds appear in discussions of holography~\cite{de2016entanglement,bousso2002holographic}. Our results concerning the edge area could be useful in bounding the covariant entropy in some causal diamond shaped regions, when the curvature scale is larger than the diamond size. Moreover, by considering the quantum speed limit of quantum operations inside a causal diamond, Seth Lloyd is able to derive the Einstein field equation~\cite{lloyd2012quantum}. It would be interesting to see how one can generalise his arguments to higher order in a vacuum causal diamond, and then possibly connect with the geometry results given in this work. Lastly, recall that we define the causal diamond as the domain of dependence of the edge, generalising the standard Alexandrov interval causal diamond definition. We believe our definition is a more natural way to understand causal diamonds as it coincides with the notion of causally closed set in algebraic quantum field theory~\cite{casini2003geometrical,haag1996local}, and resembles the entanglement wedge in AdS/CFT~\cite{rangamani2017holographic}. Although the actual diamond is not the main object of interest in our work except for ACD, we believe our definition could be potentially useful in other applications.

\section*{Acknowledgements}
We would like to thank Leo Brewin for providing some detailed expressions of RNC expansions, Ted Jacobson for pointing out the possibility of solving lightcone cut area using the Raychauduri equation, Jos\'e M.M.Senovilla for clarifying superenergy, suggesting diamond size ambiguity and possible connections with quasilocal mass, and many other useful comments and suggestions from above people. We also would like to thank Inegmar Bengtsson, Gary Gibbons, Renato Renner, Ernest Tan and Henrik Wilming for comments and discussions. This work received support from the Swiss National Science Foundation via the National Center for Competence in Research ``QSIT''.

\bibliographystyle{apsrev4-1}
\bibliography{causaldiamonds}

\end{document}